\documentclass[%
 reprint,
superscriptaddress,
 amsmath,amssymb,
 aps,
]{revtex4-2}
\usepackage{booktabs}
\usepackage{xcolor}
\usepackage{graphicx}
\usepackage{dcolumn}
\usepackage{subcaption}
\usepackage{bm}
\usepackage{multirow}
\usepackage{siunitx}
\DeclareSIUnit\dBm{\deci\bel\milli\watt}
\DeclareSIUnit\dB{\deci\bel}
\usepackage{mhchem}
\usepackage{harpoon}
\usepackage{hyperref}
\usepackage{url}
\begin{document}

\title{Millisecond phonon lifetimes in piezoelectric nanomechanical resonators}

\author{Oliver~A.~Hitchcock}
\thanks{These authors contributed equally}
 \affiliation{\mbox{Department of Applied Physics and Ginzton Laboratory, Stanford University, Stanford, California 94305, USA}}
 \affiliation{Department of Physics, Stanford University, Stanford, CA 94305, USA}
\author{Matthew~P.~Maksymowych}
\thanks{These authors contributed equally}
 \affiliation{\mbox{Department of Applied Physics and Ginzton Laboratory, Stanford University, Stanford, California 94305, USA}}
\author{Kaveh~Pezeshki}
\thanks{These authors contributed equally}
 \affiliation{\mbox{Department of Applied Physics and Ginzton Laboratory, Stanford University, Stanford, California 94305, USA}}
\author{Takuma~Makihara}
 \affiliation{\mbox{Department of Applied Physics and Ginzton Laboratory, Stanford University, Stanford, California 94305, USA}}
\author{Erik~Szakiel}
 \affiliation{\mbox{Department of Applied Physics and Ginzton Laboratory, Stanford University, Stanford, California 94305, USA}}
\author{Chou-Wei~Kiang}
 \affiliation{\mbox{Department of Applied Physics and Ginzton Laboratory, Stanford University, Stanford, California 94305, USA}}
 \affiliation{Department of Electrical Engineering, Stanford University, Stanford, CA 94305, USA}
  \author{Daniel~Rugar}
 \affiliation{\mbox{Department of Applied Physics and Ginzton Laboratory, Stanford University, Stanford, California 94305, USA}}
\author{\\Amir~H.~Safavi-Naeini}
\email{safavi@stanford.edu}
\affiliation{\mbox{Department of Applied Physics and Ginzton Laboratory, Stanford University, Stanford, California 94305, USA}}

\date{\today}

\begin{abstract}
Piezoelectric nanomechanical resonators operating at gigahertz frequencies are a promising platform for quantum information processing and precision sensing, but phonon lifetimes have been limited to the microsecond range, widely attributed to the higher intrinsic loss of piezoelectric materials. Here we show that this limitation is not fundamental. Using a fabrication process informed by recent studies of defect-mediated loss in lithium niobate, we combine chemical restoration of ion-milled surfaces with a shadow-mask metallization process to produce lithium niobate phononic crystal resonators with inferred internal phonon lifetimes of up to \SI{21}{\milli\second} at \SI{3.6}{\GHz}, corresponding to quality factors exceeding $4\times10^8$ ($f\cdot Q_i > 10^{18}$) at millikelvin temperatures. By systematically varying the electrode-to-resonator spacing, we demonstrate that the measured lifetime is set by external coupling between the device and the measurement circuit rather than material loss. This establishes that lithium niobate, once fabrication-induced surface damage is removed, achieves an intrinsic quality comparable to the best mechanical materials while remaining an intrinsically piezoelectric platform with strong electromechanical coupling.
\end{abstract}

\maketitle

\section{Introduction}

Gigahertz-frequency nanomechanical resonators are a rapidly maturing platform for quantum information processing, with applications ranging from microwave-to-optical transduction and quantum memory to precision mass sensing. For these applications, a long phonon lifetime is essential: it sets the storage time of a quantum memory, the sensitivity of a sensor, and the gate count of an acoustic processor. The longest-lived devices are fabricated from single-crystal silicon, where phonon lifetimes now exceed \SI{20}{\milli\second}~\cite{MacCabe2020,Bozkurt2025}. But silicon is not piezoelectric: coupling it to electromagnetic circuits requires electrostatic biasing or intermediate transducers, yielding effective electromechanical coupling coefficients ($k_\text{eff}^2\sim10^{-6}$) roughly a thousand times weaker than intrinsically piezoelectric materials~\cite{Bozkurt2023, Bozkurt2025}. Lithium niobate (LN) and other strong piezoelectrics can provide $k_\text{eff}^2 \sim 10^{-3}$ to $10^{-1}$, but nanoscale resonators on these platforms have been confined to microsecond lifetimes~\cite{Arrangoiz-Arriola2019,Wollack2022}, three to four orders of magnitude shorter than those measured in silicon~\cite{MacCabe2020, Bozkurt2025}. This enormous gap has led much of the community to treat piezoelectric materials as inherently lossy, instead pursuing hybrid architectures that confine the acoustic mode primarily in non-piezoelectric substrates such as silicon and sapphire while using a thin piezoelectric film only for transduction~\cite{Chu2017, Marti2024, Franse2024}. The implicit assumption has been that high material quality and strong piezoelectricity are incompatible.

Several lines of evidence have recently begun to challenge this assumption. Measurements of bulk acoustic wave resonators in LN, where surface participation is low, revealed quality factors far exceeding those of nanoscale devices from the same material~\cite{Gruenke-Freudenstein2025}, implicating surfaces rather than the bulk crystal lattice as the source of loss. Detailed materials studies identified several mechanisms and tied them to specific fabrication steps. Ion milling, for example, the standard technique for patterning chemically inert LN, amorphizes exposed surfaces and creates a ${\sim}\SI{10}{\nano\meter}$ layer hosting a dense population of two-level system (TLS) defects that couple to mechanical modes via strain~\cite{Gruenke-Freudenstein2025, Maksymowych2025,yuksel2025}. Conventional resist-based metallization introduces additional organic contamination that cannot be easily removed. In a \SI{150}{\nano\meter}-thick film, where surface participation is high, these defects dominate the energy decay. Notably, earlier attempts to improve lifetime simply by increasing the electrode-to-mode distance did not yield significant gains~\cite{Wollack2021}, because the underlying surface damage from the fabrication process remained.

\begin{figure*}[ht!]
\centering
\includegraphics[width=\textwidth]{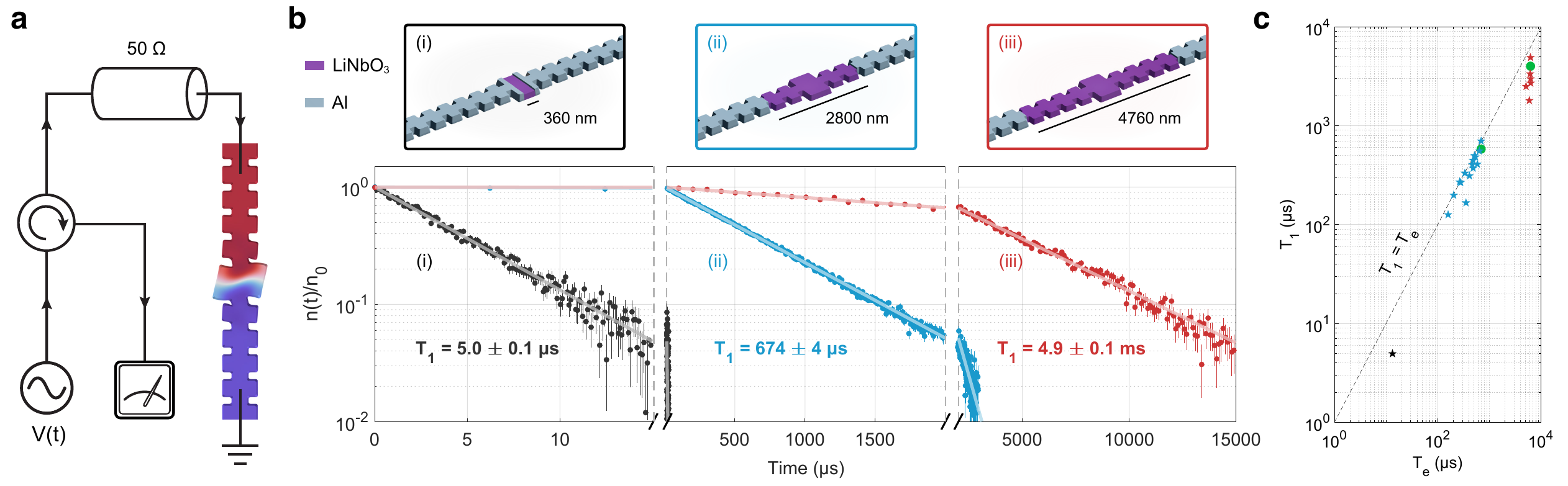}
\caption{\textbf{Device architecture and phonon lifetime measurements.} \textbf{a,}~Schematic of our RF reflectometry measurement setup. The LN phononic crystal resonator is cooled to $\SI{10}{\milli\kelvin}$ in a dilution refrigerator and coupled to a microwave transmission line via aluminum electrodes while the other electrode is grounded. The simulated mechanical shear mode deformation produces the colormapped voltage via the piezoelectric effect. The voltage profile indicates ideal electrode positions to excite the mode of interest. The electrode spacing controls the external coupling rate $\kappa_e$. After the mode is excited, it emits a voltage signal which decays proportionally with the phonon number. \textbf{b,} Device models for three electrode spacings: \SI{360}{\nano\meter}, \SI{2800}{\nano\meter}, and \SI{4760}{\nano\meter} are shown in black, blue and red frames, respectively. The corresponding phonon number ringdowns normalized with respect to starting phonon numbers of $n_0 = \{2\times 10^5, 5.3\times 10^2, 6.1\times 10^3\}$, respectively, are shown on a common x-axis, which has breaks at \SI{15}{\micro\second} and \SI{2}{\milli\second} for clarity. The exponential fits (solid lines) reveal that $T_1$ increases from \SI{5}{\micro\second} to \SI{4.9}{\milli\second} as electrode spacing increases. The stated lifetime uncertainties are 95\% confidence intervals, while data point error bars represent the standard error of the measured microwave signal power over $N=\{10^5,10^5,10^4\}$ averages.  \textbf{c,} The $T_1$ versus $T_e$ for all the resonators in the study (colored according to electrode type) with single-phonon drive power results (green circles).}
\label{fig:fig1}
\end{figure*}

Building on this understanding, we redesigned the LN nanoresonator fabrication process to address all identified sources of surface degradation. Chemical treatments that can strip ion-milling damage from LN surfaces have been developed in prior work~\cite{Gruenke-Freudenstein2025, Wollack2021}, but a central challenge has prevented their application to nanomechanical devices: the wet chemistry required for both device release and surface restoration is incompatible with metal contacts, and conventional resist-based liftoff metallization contaminates the very surfaces being cleaned. We resolve this incompatibility by introducing a shadow-mask metallization process in which aluminum is deposited through a freestanding silicon nitride stencil as the final fabrication step, after all surface cleaning is complete. 

Shadowmasks decouple surface preparation from metallization, enabling aggressive chemical treatment of the fully released structures before any metal touches the device. We then perform a direct and sensitive test of the resulting material quality. Phononic crystal resonators fabricated with this process from $\SI{150}{\nano\meter}$ thick LN, operating in the 3--\SI{4}{\GHz} range, are measured at millikelvin temperatures with various electrode spacings that set the external coupling rate. As we move the electrodes farther from the resonant mode, external coupling decreases, and the total phonon lifetime rises from \SI{5}{\micro\second} to \SI{4.9}{\milli\second}, tracking the external coupling limit at each point (Fig.~\ref{fig:fig1}c). This demonstrates that the intrinsic material loss is far below the levels previously assumed, establishing an internal quality factor $Q_i > 4\times 10^8$ ($f{\cdot} Q_i > 10^{18}$). Because LN is intrinsically piezoelectric, this material quality can in principle be accessed at any coupling strength: the electrode spacing is a design parameter, not a fundamental constraint, and coupling can be further increased by improving the metal quality or by moving the metal off the resonator entirely and coupling through near-field electromagnetic interaction. 

In this work, we demonstrate millisecond phonon lifetimes in an intrinsically piezoelectric platform and show that that the poor coherence of previous LN nanomechanical resonators was a consequence of fabrication-induced surface damage. Our results re-frame the prospects for piezoelectric nanomechanics: the material was not the problem.

\begin{figure*}[t!]
\centering
\includegraphics[width=\textwidth]{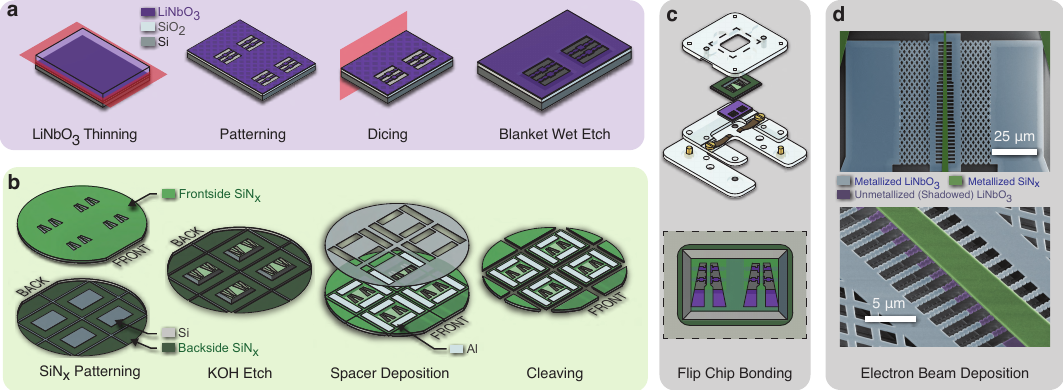}
\caption{\textbf{Shadow-mask metallization process.} \textbf{a,}~LN device fabrication: film thinning with SC1 wet etch to remove ion-mill damage, electron-beam patterning, ion-mill etch, wafer dicing, HF release and SC1 surface cleaning. \textbf{b,}~Silicon nitride shadow-mask fabrication: frontside patterning, KOH backside Si etch to release freestanding membrane, aluminum spacer deposition using a separately fabricated stencil mask (grey), and cleaving into chips. \textbf{c,}~The shadow-mask and device chip are aligned in a flip-chip bonder, clamped in a custom fixture, and loaded into an electron-beam evaporator for aluminum deposition. \textbf{d,}~Colorized SEM of a completed device prior to mask removal showing metallized LN (blue), metallized silicon nitride (green), and unmetallized (shadowed) LN in the defect region (purple). The shadow-mask prevents metal deposition on the defect cell, ensuring the resonant mode interacts only with pristine LN surfaces.}
\label{fig:fig2}
\end{figure*}

\section{Device architecture and phonon lifetime measurements}

Our devices are phononic crystal resonators fabricated from a \SI{150}{\nano\meter}-thick film of X-cut MgO-doped LN on a \SI{3}{\micro\meter} buried oxide layer with a \SI{700}{\micro\meter} high resistivity silicon handle (NanoLN). A quasi-one-dimensional array of periodic cells supports a complete acoustic bandgap, with a larger defect cell ($\sim 600 \times \SI{900}{\nano\meter\squared}$) at the center hosting a shear-mode mechanical resonance in the 3--\SI{4}{\GHz} range (Fig.~\ref{fig:fig1}a). Aluminum electrodes deposited via a shadow-mask form contacts to a microwave transmission line, enabling piezoelectric excitation and readout of the mechanical mode via standard reflectometry~\cite{Wollack2021,yuksel2025,Maksymowych2025}.

The critical design parameter is the electrode-to-defect spacing: the distance between the metal contacts and the central localized resonant mode. This spacing controls the external coupling rate $\kappa_e$ while leaving the intrinsic material properties unchanged. By fabricating devices with three different electrode spacings (\SI{360}{\nano\meter}, \SI{2800}{\nano\meter}, and \SI{4760}{\nano\meter}), we tune $\kappa_e$ over more than three orders of magnitude on chips processed with the same fabrication protocol.

We cool devices to \SI{10}{\milli\kelvin} in a dilution refrigerator and characterize them using microwave reflectometry (Appendices~\ref{app:measurement_setup} and \ref{app:spectroscopy}). We extract the external coupling rate $\kappa_e$ from the depth of the reflection dip with vector network analyzer (VNA) spectroscopy. We measure the phonon lifetime $T_1 = 1/\kappa$ by observing the energy ringdown of coherent states, calibrating the intracavity phonon number using $\kappa_e$ and the known input power (Appendix~\ref{app:phonon_number}).

\begin{table}[b]
\centering
\caption{Measured and inferred device parameters for three electrode spacings. $T_e\equiv 1/\kappa_e$ is the measured external coupling lifetime inferred from VNA spectroscopy. $T_1$ is the measured total phonon lifetime from ringdown at \SI{10}{\milli\kelvin}. $T_{1,\mathrm{in}}$ is the inferred internal loss ringdown time. For \SI{2800}{\nano\meter}, the lifetime is consistent with the external coupling limit, therefore we report a one-sided bound.}
\label{tab:devices}
\begin{ruledtabular}\renewcommand{\arraystretch}{1.1}
\begin{tabular}{c c cl}
Electrode spacing& $T_e = 1/\kappa_e$ & $T_1$  &$T_{1,\mathrm{in}}$\\
\hline
\SI{360}{\nano\meter} & $\SI{13.30(9)}{\micro\second}$ & $\SI{5.0(1)}{\micro\second}$  & \SI{8.0\pm.3}{\micro\second}\\
\SI{2800}{\nano\meter} & $\SI{690(30)}{\micro\second}$ & $\SI{674(4)}{\micro\second}$  &$>$\SI{10}{\milli\second}\\
\SI{4760}{\nano\meter} & $\SI{6.4(5)}{\milli\second}$ & $\SI{4.9 \pm 0.1}{\milli\second}$  & \SI{21\pm6}{\milli\second}\\
\end{tabular}
\end{ruledtabular}
\end{table}

The results are summarized in Table~\ref{tab:devices} and Fig.~\ref{fig:fig1}. As the electrode spacing increases from \SI{360}{\nano\meter} to \SI{4760}{\nano\meter}, the external coupling rate $\kappa_e/2\pi$ drops from $\SI{11.97(8)}{\kHz}$ to $\SI{25(2)}{\Hz}$ and the measured phonon lifetime rises correspondingly from \SI{5.0(1)}{\micro\second} to \SI{4.9(1)}{\milli\second}, spanning three orders of magnitude (Fig.~\ref{fig:fig1}c).

The key observation is the relationship between total lifetime $T_1$ and external coupling lifetime $T_e\equiv1/\kappa_e$ at each spacing. For the two widest spacings, the measured $T_1$ is either consistent with or close to the external coupling limit $T_e$, indicating that the energy decay is dominated by leakage into the measurement circuit rather than by intrinsic material loss. For the \SI{4760}{\nano\meter} device, $T_1 = \SI{4.9\pm.1}{\milli\second}$ compared to $T_e = \SI{6.4\pm.5}{\milli\second}$ yields an internal lifetime of $T_{1,\text{in}} =\SI{21\pm6}{\milli\second}$ and a corresponding $Q_i = \SI{5\pm1e8}{}$. 

For the closest electrode spacing (\SI{360}{\nano\meter}), $T_1 = \SI{5.0\pm.1}{\micro\second}$ is significantly shorter than $T_e = \SI{13.30(9)}{\micro\second}$, indicating the presence of an additional loss channel. The implied internal lifetime of $T_{1,\text{in}} =\SI{8.0\pm.3}{\micro\second}$ for this device likely reflects loss associated with the metal--LN interface, as the electrodes are in close proximity to the resonant mode. This observation is consistent with the systematic trend: as the metal is moved away, the excess loss vanishes and the device becomes external-coupling limited. Fig. \ref{fig:fig1}c depicts the $T_1$ versus $T_e$ for all resonators in the study (stars) including data taken with single-phonon drive powers (green circles). The plot shows that external-coupling dominance and millisecond phonon lifetimes across multiple devices are retained even at single-phonon occupation levels. We attribute the long $T_1$ at single-phonon drive powers to a low number of intrinsic defects in these devices.
The reproducibility of these results across devices, designs, and deliberate
variations of the fabrication flow are detailed in
Appendix~\ref{app:fab_variants}.

These measurements demonstrate that the internal phonon lifetime $T_{1,\text{in}}$ of our LN resonators exceeds \SI{10}{\milli\second} at gigahertz frequencies and millikelvin temperatures, corresponding to $Q_i > 4\times 10^8$ and a frequency--quality factor product $f{\cdot} Q_i > 10^{18}$. This is comparable to the best silicon nanomechanical resonators~\cite{Bozkurt2025} and exceeds all previous piezoelectric nanomechanical devices by more than three orders of magnitude.

\section{Shadow-Mask Metallization}

A central challenge in fabricating high-coherence LN nanomechanical resonators is the incompatibility between surface preparation and metallization. Acid-base treatments can strip the amorphous damage layer left by ion milling and remove organic residue from polymer resist processing~\cite{Gruenke-Freudenstein2025, Wollack2021}, but these treatments generally attack metals and so cannot be used after metallization. Furthermore, conventional resist-based liftoff must be performed before device release. To accommodate these limitations previous devices were fabricated with surfaces that carried both the damage introduced during patterning and residual polymer contamination from the liftoff process~\cite{Arrangoiz-Arriola2019, Wollack2022, Lee2023, Cleland2024, Maksymowych2025}.

Our shadow-mask process breaks this ordering constraint (Fig.~\ref{fig:fig2}). We fabricate freestanding silicon nitride membrane stencils, each patterned with apertures corresponding to the electrode geometry for all devices on a target chip. The process achieves sub-micron feature sizes suitable for the $\ge \SI{360}{\nano\meter}$ electrode spacings of our devices via freestanding membrane bridges (Fig.~\ref{fig:fig2}b, Appendix~\ref{app:shadowmask_apparatus}). The LN phononic crystals are patterned by ion milling, released in hydrofluoric acid, and subjected to a full course of $\text{NH}_4\text{OH} - \text{H}_2\text{O}_2 - \text{H}_2\text{O}$ (SC1) base cleaning to remove the amorphous damage layer (Fig. \ref{fig:fig2}a). This SC1 clean follows the release when all surfaces of the phononic crystal are accessible. A shadow-mask is then aligned to the device chip in a custom jig fixture using a commercial flip-chip bonder (Fig. \ref{fig:fig2}c). Aluminum is deposited by electron-beam evaporation through the stencil apertures as the final fabrication step, forming the device electrodes (Fig.~\ref{fig:fig2}d). No photoresist or organic material ever contacts the cleaned device surfaces, and no additional processing is done following metallization and prior to measurement.

This approach offers several advantages beyond surface cleanliness. Because the shadow-mask relies only on mechanical alignment and physical vapor deposition, it is not specific to LN and can be applied to any substrate material where surface quality limits device performance. The shadow-mask can be designed to produce a tapered electrode geometry across a resonator array, allowing a single deposition to metallize devices with a range of electrode spacings on one chip. The shadow-mask is reusable and the process is compatible with post-release annealing. Furthermore, this technique is extensible to very large patterns. Characterization of the mask-to-device alignment shows no evidence of systematic membrane distortion across a \qty[parse-numbers=false]{6.5\times4.5}{\milli\meter\squared} membrane, with alignment offsets consistent with the mechanical tolerance of the bonding tool (Appendix~\ref{app:shadowmask_distortion}).

\begin{figure}[ht!]
    \centering
    \includegraphics[width=1.0\columnwidth]{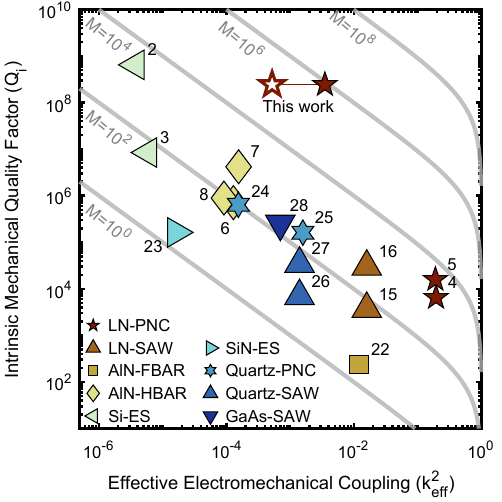}
    \caption{\textbf{Comparison of internal quality factor versus effective electromechanical coupling}. Parameters are shown for this work and various other acoustic materials and platforms from the literature, with each point label corresponding to the reference number. Literature $k_{\text{eff}}^2$ values are simulation- or circuit-model-derived (Appendix~\ref{app:k2_estimation}); for this work, the bar spans the parasitics-loaded measurement (unfilled) and the simulated intrinsic value (filled). Gray lines denote constant resonator figure of merit: $M = k_{\text{eff}}^2 Q_i / (1 - k_{\text{eff}}^2)$.}
    \label{fig:Fig_compare_performance}
\end{figure}

\section{Performance in Context}

To compare our devices against the diverse landscape of mechanical resonator platforms with different materials, geometries, frequencies, and coupling schemes, we plot the internal quality factor $Q_i$ at single-phonon occupations against the estimated effective electromechanical coupling coefficient $k_{\text{eff}}^2$ (Fig.~\ref{fig:Fig_compare_performance}). These two dimensionless, intensive parameters capture the essential physics: $Q_i = \omega_0 T_{1,\text{in}}$ measures energy storage independent of frequency, while $k_{\text{eff}}^2$ measures the efficiency of energy transduction between electrical and mechanical domains independent of mode volume and circuit design (Appendix~\ref{app:k2_estimation}).

The resulting landscape reveals a striking pattern. All previous devices fall along an approximate tradeoff: platforms with strong electromechanical coupling (LN phononic crystals~\cite{Wollack2022, Arrangoiz-Arriola2019}, LN SAWs~\cite{Satzinger2018, Bienfait2019}) cluster at $Q_i \sim 10^3$--$10^4$, while platforms with high $Q_i$ (silicon electrostatic~\cite{Bozkurt2025, Bozkurt2023}, AlN HBARs~\cite{Marti2024, Chu2017, Franse2024}) achieve coherence only at $k_{\text{eff}}^2 \lesssim  10^{-4}$. Our new devices break this pattern by occupying a previously inaccessible region of parameter space; one that simultaneously has $Q_i > 10^8$ and $k_{\text{eff}}^2 = \qtyrange{5.2e-4}{3.5e-3}{}$ inferred from measurements and simulations, as detailed in Appendix~\ref{app:k2_estimation}. 

These two parameters can be combined into a resonator figure of merit~\cite{IEEE1988},
\begin{equation}\label{eq:definition_resonator_figure_of_merit_M}
    M = \frac{k_{\text{eff}}^2}{1 - k_{\text{eff}}^2} Q_i \approx k_{\text{eff}}^2 Q_i,
\end{equation}
which in the language of circuit quantum electrodynamics scales as $M \sim g^2 / \omega\kappa$ and captures the overall utility of an electromechanical device, with $g$ representing circuit-mechanics coupling and $\omega$ and $\kappa$ representing the mechanical frequency and energy decay. Our devices achieve a figure of merit between $\SI{1.3e5}{}$ and $\SI{8.6e5}{}$, exceeding all previous platforms by more than an order of magnitude (Table~\ref{tab:table_literature_parameter}).

\section{Conclusions}

Our results demonstrate that the quality of LN as a material for nanomechanics is comparable to that of the best non-piezoelectric platforms, once fabrication-induced surface damage is removed. The historical coupling--coherence tradeoff that has shaped the field was not a fundamental constraint but a consequence of how these devices were made. Because our devices remain limited by external coupling rather than intrinsic material loss, the true quality of the material has not yet been reached. Moreover, congruently grown LN is an extremely defective crystal, with approximately one niobium antisite defect per cubic nanometer~\cite{Volk2008}; as surface damage is eliminated, bulk defect engineering becomes the next frontier, and we anticipate that orders of magnitude of improvement in $T_1$ remain accessible.

Several directions for future work follow from these results. The phononic crystal geometry itself may enable engineering of TLS dynamics: recent work has shown that the phononic bandgap can modify defect relaxation rates by reshaping the local phonon density of states~\cite{Chen2024}, while the strong coupling between individual TLS and the mechanical mode~\cite{yuksel2025} provides a direct spectroscopic handle on the defects that limit coherence. Beyond quantum information, these resonators are natural candidates for precision mass sensing~\cite{Hanay2012}, where the ultimate sensitivity is limited by frequency noise from TLS~\cite{Maksymowych2025, Hitchcock2026}. The dramatically reduced defect density in our devices may make this noise manageable for the first time. The immediate next step is to achieve strong coupling to superconducting qubits and use the ultra-long $T_1$ times demonstrated here to enter new regimes for quantum acoustic systems.

More broadly, because the shadow-mask metallization process is material-agnostic, the same fabrication paradigm can be applied to other piezoelectric platforms such as quartz, gallium arsenide, and aluminum nitride, wherever surface damage currently limits device performance. However, LN is unique for several reasons. It offers high material quality, strong intrinsic piezoelectricity, and is compatible with scalable nanofabrication. Furthermore, unlike silicon whose surfaces oxidize rapidly upon atmospheric exposure, LN is itself an oxide and its surfaces are inherently stable in air. Our devices were exposed to atmosphere for extended periods between cooldowns with no additional cleaning after the shadow-mask metallization, yet retained the long lifetimes reported here. These factors position LN nanomechanical resonators as a practical platform for quantum memory, multimode acoustic processing, and precision sensing at the quantum limit.


\begin{acknowledgments}
The authors thank Felix Mayor, Sultan Malik, and John Mamin for useful discussions and Ali Khalatpour and Rachel Gruenke-Freudenstein for collaborating to develop the SC1 chemical process. This work was funded by Amazon Web Services Inc., the Air Force Office of Scientific Research and the Office of Naval Research under award number FA9550-23-1-0338, the Gordon and Betty Moore Foundation (Grant 12214), a Moore Inventor Fellowship, the U.S.\ Department of Energy through grant no.\ DE-AC02-76SF00515 and via the Q-NEXT Center. This work was supported in part by NIH grant RF1MH136394. We thank IARPA and MIT Lincoln Labs for providing us with the TWPA. Device fabrication was performed at the Stanford Nano Shared Facilities (SNSF) and the Stanford Nanofabrication Facility (SNF), supported by NSF award ECCS-2026822. This material is based upon work supported by the Air Force Office of Scientific Research under award number FA9550-23-F-0014 in the amount of \$139,400. M.P.M. acknowledges support from a Natural Sciences and Engineering Research Council of Canada (NSERC) PGS-D fellowship. K.P. acknowledges support from the Department of Defense (DoD) through the National Defense Science \& Engineering Graduate (NDSEG) Fellowship Program. C.-W.K. acknowledges support from National Science Foundation Graduate Research Fellowship Program under grant No. DGE-2146755.
\end{acknowledgments}

\bibliography{references}

\onecolumngrid
\appendix
\newpage
\section{Measurement Setup} \label{app:measurement_setup}

\begin{figure}[ht]
	\centering
	\includegraphics[width=0.5\columnwidth]{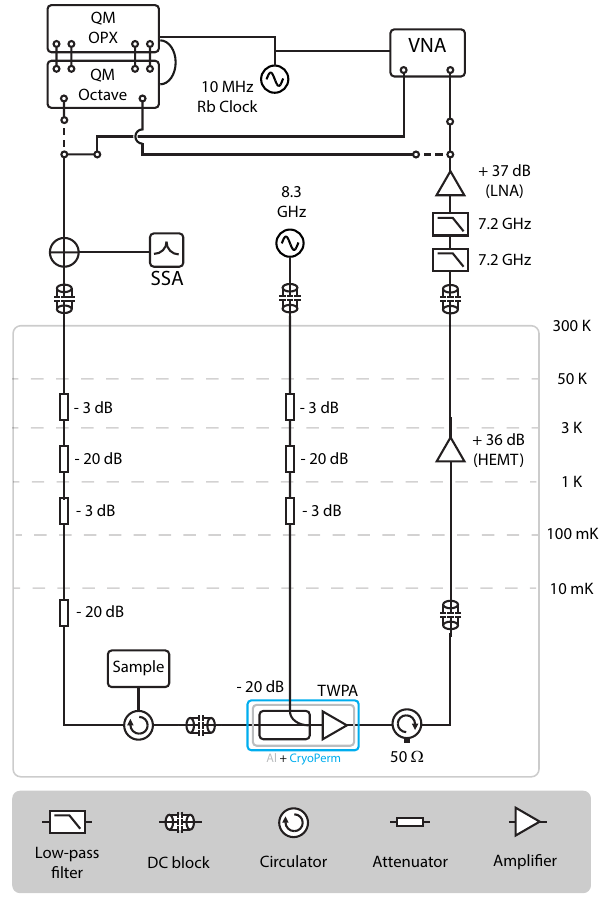}
	\caption{\textbf{Microwave measurement setup.} Schematic of the cryogenic wiring and room-temperature electronics used for reflectometry measurements. The device is thermalized to \SI{10}{\milli\kelvin} at the base stage of a dilution refrigerator. The input line is attenuated to suppress thermal noise, and the output signal is amplified by a near-quantum-limited TWPA followed by a HEMT and room-temperature LNA. A VNA is used for spectroscopy and a Quantum Machines OPX for pulsed ringdown measurements.}
	\label{fig:measurement_setup}
\end{figure}

The microwave measurement setup is shown in Fig.~\ref{fig:measurement_setup}. Devices are mounted at the base stage of a dilution refrigerator at \SI{10}{\milli\kelvin} and measured in a one-port reflection geometry. The input line carries a total of \SI{46}{\dB} of cryogenic attenuation distributed across the \SI{50}{\kelvin}, \SI{3}{\kelvin}, \SI{1}{\kelvin} and \SI{10}{\milli\kelvin} stages to thermalize the microwave noise to the base temperature.
Microwave signals are routed to and from the device through a circulator chain at the base stage. The output signal is amplified by an aluminum traveling-wave parametric amplifier (TWPA) housed in a combined aluminum and CryoPerm magnetic shield at the base stage, providing ${\sim}\SI{16}{\dB}$ of near-quantum-limited gain, followed by a high-electron-mobility transistor (HEMT) amplifier ($+\SI{36}{\dB}$) at \SI{3}{\kelvin} and a room-temperature low-noise amplifier (LNA, $+\SI{37}{\dB}$). The TWPA pump tone at \SI{8.3}{\GHz} is injected through a dedicated line with \SI{26}{\dB} of nominal attenuation, and lowpass filters at \SI{7.2}{\GHz} on the output line reject the pump before the room-temperature electronics. Two measurement instruments are used: a vector network analyzer (VNA) for continuous-wave spectroscopy and a Quantum Machines OPX with Octave upconversion module for pulsed ringdown measurements. Both are referenced to a common \SI{10}{\MHz} rubidium clock. A signal source analyzer (SSA) monitors the phase noise of the local oscillator and spurious sidebands.

\section{Extraction of $\kappa_e$ from VNA Spectroscopy} \label{app:spectroscopy}

The resonant modes are characterized via microwave reflectometry using a vector network analyzer with an effective measurement bandwidth of \SI{1}{\hertz}. We fit each measured reflection spectrum to the single-mode model
\begin{equation}
    \langle S_{11}(\Delta) \rangle = S_{\text{bg}}\left(1 - e^{i\phi}\frac{\kappa_e}{i\Delta + \kappa_\text{tot}/2}\right), \label{eq:S11_fit}
\end{equation}
where $\Delta$ is the drive detuning from the mode frequency, $\kappa_e$ is the external coupling rate, $\kappa_\text{tot}$ is the total spectroscopic linewidth, $\phi$ is an asymmetry parameter accounting for impedance mismatch, $S_\text{bg}$ describes the microwave background and $\langle\cdot\rangle$ denotes the coherent average over repeated frequency sweeps. On resonance ($\Delta = 0$), the reflection coefficient reduces to $S_{11}(0) = S_\text{bg}(1 - e^{i\phi} \cdot 2\kappa_e/\kappa_\text{tot})$: the dip depth is governed by the ratio $\kappa_e/\kappa_\text{tot}$, while the linewidth gives $\kappa_\text{tot}$. Together, these determine $\kappa_e$.

\begin{figure}[ht]
	\centering
	\includegraphics[width=0.5\columnwidth]{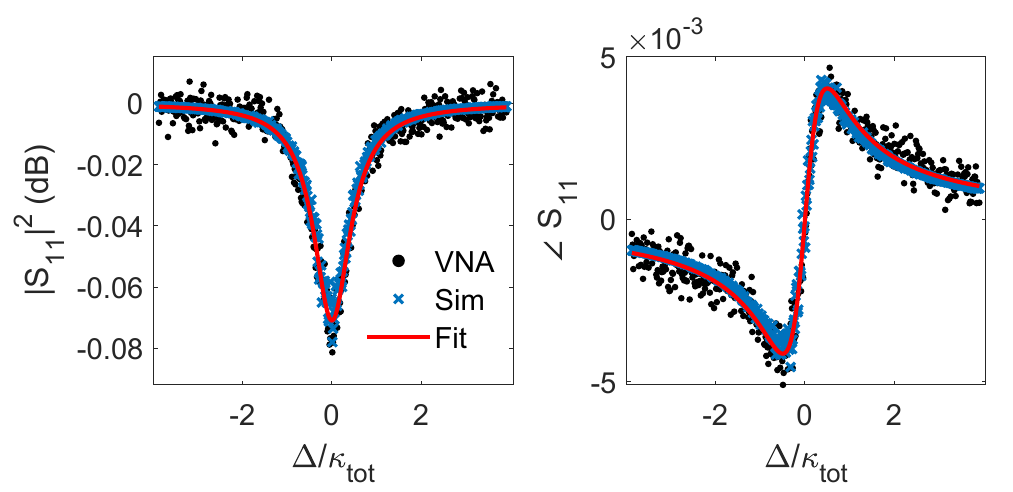}
	\caption{\textbf{Microwave Reflectometry Measurement.} The measured magnitude and phase response of a $\SI{3.35}{\GHz}$  device (at \SI{10}{\milli\kelvin} and $n\approx 100$~phonons) with a designed electrode spacing of \SI{2800}{\nano\meter} is shown as black dots. The corresponding fit of experimental data to Eq.~\ref{eq:S11_fit} is shown as a red line, while the simulated reflection spectrum from our noisy Langevin model is shown as blue x's.}
	\label{fig:S11_meas}
\end{figure}

In our devices, the spectroscopic linewidth is dominated by frequency noise (dephasing) rather than energy decay. For the \SI{2800}{\nano\meter}-spacing device data shown in Fig.~\ref{fig:S11_meas} (black dots), the fit to Eq.~\ref{eq:S11_fit} (red line) yields $\kappa_\text{tot}/2\pi \approx \SI{60}{\kilo\hertz}$, while the energy decay rate in the ringdown is $\kappa/2\pi \approx \SI{236}{\hertz}$. The total linewidth can be decomposed as $\kappa_\text{tot} = \kappa + 2\Gamma_\phi$, where $\Gamma_\phi$ is the dephasing rate and $\kappa = \kappa_e + \kappa_i$ is the energy decay rate. When the frequency noise has a correlation time $\tau_c$ short compared to the cavity lifetime $1/\kappa$ and dephasing time $T_\phi=1/\Gamma_\phi$, the slow time-averaged VNA response retains the Lorentzian form of Eq.~\ref{eq:S11_fit} with $\kappa_\text{tot} = \kappa + 2\Gamma_\phi$, and $\kappa_e$ is extracted correctly from the fit even though the apparent linewidth is inflated by dephasing. The Lorentzian quality of our measured spectra is consistent with this regime. For a detailed derivation of the VNA response in the presence of frequency noise, see Ref.~\cite{ Maksymowych2025}. 
The values of $\kappa_e$ reported in Table~\ref{tab:devices} are obtained by averaging this fit over multiple VNA sweeps for each device.

To verify this assertion numerically, we simulate the stochastic Langevin equation describing the resonator driven through a frequency noise process~\cite{Maksymowych2025},  compute the time-averaged $S_{11}$ spectrum, and fit it to Eq.~\ref{eq:S11_fit}. 
 We find that the fit estimate of $\kappa_e$ agrees with the true value of $\kappa_e$ across a wide range of dephasing rates at least up to $\Gamma_\phi/\kappa\approx10^4$ which is significantly more dephasing than our devices exhibit. In Fig.~\ref{fig:S11_meas} the simulated average reflection spectrum with $\kappa/2\pi=\kappa_e/2\pi=\SI{236}{\hertz}$, $\Gamma_\phi/2\pi = \SI{30}{\kilo\hertz}$, and $\tau_c^{-1}\gg \{\Gamma_\phi,\kappa\}$ is depicted as blue x's, showing close overlap with the experimental data.  

\begin{table*}[hb!]
\centering
\caption{Extended device parameters for the three electrode spacings studied in Fig.\ref{fig:fig1}. Simulated values of $\kappa_e$ are obtained from finite-element modeling of the phononic crystal geometry. Measured $\kappa_e$ and $\kappa_\text{tot}$ are extracted from VNA spectroscopy (Eq.~\ref{eq:S11_fit}). The total energy decay rate $\kappa = 1/T_1$ is obtained from ringdown measurements. The internal loss rate $\kappa_i = \kappa - \kappa_e$ and dephasing rate $\Gamma_\phi = (\kappa_\text{tot} - \kappa)/2$ are inferred from these quantities. Inferred intrinsic losses for the \SI{2800}{\nano\meter} device are reported as one-sided bounds due to large relative uncertainties.}
\label{tab:extended_devices}
\renewcommand{\arraystretch}{1.4}
\begin{ruledtabular}
\begin{tabular}{lccc}
 & \textbf{\SI{360}{\nano\meter} spacing} & \textbf{\SI{2800}{\nano\meter} spacing} & \textbf{\SI{4760}{\nano\meter} spacing} \\
\hline
\multicolumn{4}{l}{\textit{\textbf{Design parameters}}} \\
Defect cell (nm $\times$ nm) & $700\times1100$ & $600\times1000$ & $550\times950$ \\
Measured frequency & \SI{4.20}{\GHz} & \SI{3.35}{\GHz} & \SI{3.60}{\GHz} \\[4pt]
\multicolumn{4}{l}{\textit{\textbf{Simulated coupling}}} \\
$\kappa_e^\text{sim}/2\pi$ & \SI{10.8}{\kilo\hertz} & \SI{330}{\hertz} & \SI{110}{\hertz} \\
$T_e^\text{sim} = 1/\kappa_e^\text{sim}$ & \SI{15}{\micro\second} & \SI{500}{\micro\second} & \SI{1.4}{\milli\second} \\[4pt]
\multicolumn{4}{l}{\textit{\textbf{Measured coupling (VNA spectroscopy)}}} \\
$\kappa_e/2\pi$ & $\SI{11.97(8)}{\kilo\hertz}$ & $\SI{230(10)}{\hertz}$ & \SI{25(2)}{\hertz} \\
$T_e = 1/\kappa_e$ & $\SI{13.30 \pm 0.09}{\micro\second}$ & $\SI{690 \pm 30}{\micro\second}$ & $\SI{6.4(5)}{\milli\second}$ \\
$\kappa_\text{tot}/2\pi$ & $\SI{147 \pm 1}{\kilo\hertz}$  & $\SI{60 \pm 3}{\kilo\hertz}$ & $\SI{34 \pm 3}{\kilo\hertz}$ \\[4pt]
\multicolumn{4}{l}{\textit{\textbf{Measured lifetime (ringdown)}}} \\
$T_1$ & $\SI{5.0 \pm 0.1}{\micro\second}$ & $\SI{674(4)}{\micro\second}$ & $\SI{4.9 \pm 0.1}{\milli\second}$ \\
$\kappa/2\pi = 1/(2\pi T_1)$ & \SI{31.8(6)}{\kilo\hertz} & \SI{236(1)}{\hertz} & \SI{32.5(7)}{\hertz} \\[4pt]
\multicolumn{4}{l}{\textit{\textbf{Inferred intrinsic parameters}}} \\
$\kappa_i/2\pi = (\kappa - \kappa_e)/2\pi$ & \SI{19.9(6)}{\kilo\hertz} & $<$\SI{16}{\Hz} & \SI{7(2)}{\hertz} \\
$T_{1,\text{in}} = 1/\kappa_i$ & \SI{8.0(3)}{\micro\second} & $>$\SI{10}{\milli\second} & \SI{21(6)}{\milli\second} \\
$Q_i = \omega_0 T_{1,\text{in}}$ & $211(7) \times 10^3$ & $>$\SI{2.1e8}{} & 5(1)$\times 10^8$ \\
$\Gamma_\phi/2\pi = (\kappa_\text{tot} - \kappa)/(4\pi)$ & $\SI{58(1)}{\kilo\hertz}$ & $\SI{30(2)}{\kilo\hertz}$ & $\SI{17(2)}{\kilo\hertz}$ \\
\end{tabular}
\end{ruledtabular}
\end{table*}
\newpage

\section{Phonon Number Calibration} \label{app:phonon_number}

The intracavity phonon number $n(t)$ shown in the ringdown measurements is inferred from the demodulated signal $S(t) = \langle I(t)^2 + Q(t)^2 \rangle$ averaged over many experimental shots, which is proportional to the output photon flux:
\begin{equation}
    S(t) = G\, \langle a_\text{out}^\dagger(t)\, a_\text{out}(t) \rangle,
\end{equation}
where $G$ is the total measurement gain of the output chain and $a_\text{out} = a_\text{in} - \sqrt{\kappa_e}\, a$ is the output field expressed in terms of the intracavity $a$ and input $a_{in}$ fields. During the drive pulse, the output is dominated by the reflected drive tone, so $S_\text{ON} \approx G\, \langle a_\text{in}^\dagger a_\text{in} \rangle$. After the drive is turned off ($a_\text{in} \to 0$), only the cavity emission remains: $S_\text{OFF}(t) = G\, \kappa_e\, n(t)$, where $n(t) = \langle a^\dagger(t)\, a(t) \rangle$. Taking the ratio of these two signals eliminates the unknown measurement gain and yields the phonon number directly:
\begin{equation}
    n(t) = \frac{P_\text{in}}{\hbar\omega_0 \kappa_e} \cdot \frac{S_\text{OFF}(t) - N^2}{S_\text{ON} - N^2}, \label{eq:ratio_method}
\end{equation}
where $P_\text{in} = \hbar\omega_0 \langle a_\text{in}^\dagger a_\text{in} \rangle$ is the input power at the device and $N^2$ is the measured noise floor. This calibration requires only the known input power and the spectroscopically determined $\kappa_e$, with no assumptions about the cavity dynamics, frequency noise model, or nonlinearity. The key requirement is that the system remain in the regime where the drive is mostly reflected ($S_\text{ON} \gg S_\text{OFF}$), which is satisfied in our devices due to the large ratio of dephasing to energy decay.

Note that while dephasing broadens the spectroscopic linewidth, it does not affect the energy decay rate: the drive populates the mode at a rate set by the broadened density of states $\kappa_\text{tot}$, while energy leaks out at the physical rate $\kappa$. For devices with substantial dephasing ($\kappa_\text{tot} \gg \kappa$), the steady-state phonon number under continuous drive can exceed estimates based on the spectroscopic linewidth $n_{spec} = P_{in}/\hbar\omega_0 \times 4\kappa_e/\kappa_{\text{tot}}^2$ by a factor of $\kappa_\text{tot}/\kappa$~\cite{Maksymowych2025}. This correction is automatically accounted for in the ratio method of Eq.~\ref{eq:ratio_method}.

\section{Compilation of device parameters from literature} \label{app:literature_compilation}
\begin{figure}[hb!]
    \centering
    \includegraphics[width=.899\columnwidth]{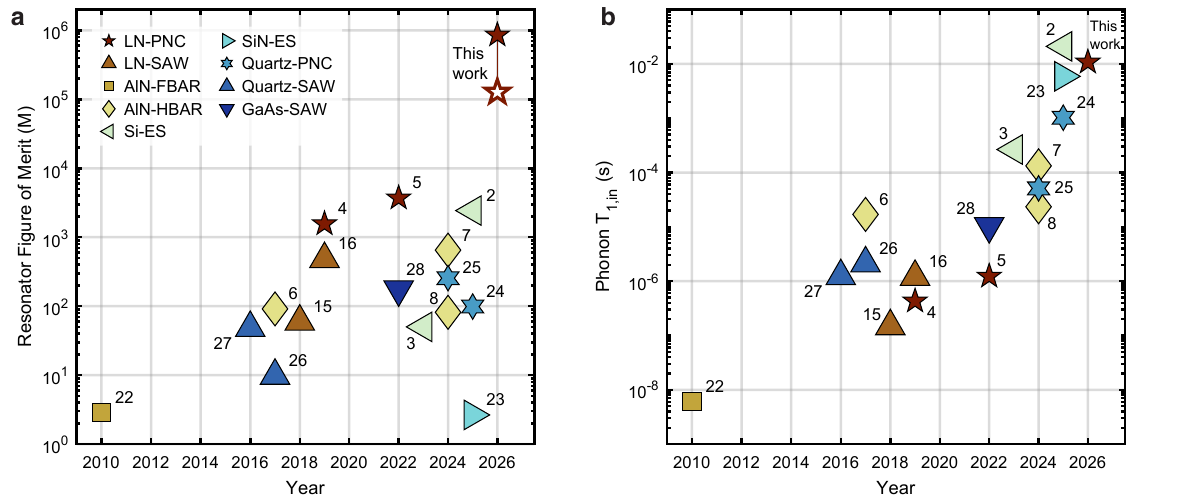}
    \caption{\textbf{Mechanical resonator performance over time.} \textbf{a,} The electromechanical resonator figure of merit $M = k_{\text{eff}}^2 Q_i / (1 - k_{\text{eff}}^2)$ and \textbf{b,} internal phonon lifetime, plotted by publication year, for devices listed in Table~\ref{tab:table_literature_parameter}. The device presented in this work achieves a figure of merit $M=\qtyrange{1.3e5}{8.6e5}{}$, a more than 50x improvement over all previous platforms, and $T_{1,\text{in}}=\SI{11}{\milli\second}$ at single-phonon drive powers, exceeding the lifetimes of previous piezoelectric devices by nearly 100x, obtaining values comparable to state-of-the-art Si and SiN electrostatic devices.}
    \label{fig:SI_M_T1_vs_year}
\end{figure}

In this section, we compile parameters from the literature for a number of electromechanical resonators and associated electromagnetic circuits into Tab.~\ref{tab:table_literature_parameter}, to facilitate comparisons of device performance across various material platforms and device types. In Fig.~\ref{fig:SI_M_T1_vs_year}a, we show the progression of the resonator figure of merit $M$ within the field over time. We find a mixed trend in which devices have been limited to $M<\SI{1e4}{}$, representing the historical tradeoff between electromagnetic coupling and phonon lifetime. Our new devices improve upon this figure of merit by over two orders of magnitude relative to previous results, indicating that this tradeoff was not fundamental. In Fig.~\ref{fig:SI_M_T1_vs_year}b, we show the progression of phonon lifetime within the field over time. This reveals a consistent upward trend that is largely the result of devices switching away from piezoelectrics as the phonon storage medium. Our current work disrupts this trend by demonstrating long phonon lifetimes with an entirely piezoelectric medium. 
\begin{table*}[h!]
\centering
\caption{Compilation of nanomechanical and quantum acoustic device parameters from literature. Parameters are quoted directly from the source when explicitly stated, or estimated from given parameters. Low-power, $\langle n\rangle\sim 1$, values for internal phonon lifetimes and quality factors are used where available. Estimation of the effective electromechanical coupling is detailed in Appendix~\ref{app:k2_estimation}.}
\label{tab:table_literature_parameter}
\renewcommand{\arraystretch}{1.5}
\resizebox{\textwidth}{!}{
\begin{tabular}{|c c c|c||c|c|c|c|c|c||c|c|c||c|c|c|}
\hline
\multicolumn{3}{|c|}{\multirow{2}{*}{\textbf{Reference}}}  & \multirow{2}{*}{\textbf{Platform}} & \multicolumn{6}{c||}{\textbf{Mechanics}}  & \multicolumn{3}{c||}{\textbf{Circuit}} & \multicolumn{3}{c|}{\textbf{Coupling}} \\ \cline{5-16} \relax
& & & & $\omega_m/2\pi$ (Hz) & {$T_{1,\text{in}}$ (s)} & {$Q_{\text{i}}$} & {$\kappa_{e}/2\pi$ (Hz)}  & {$k_{\text{eff}}^2$} & {$M$} & {$\omega_q/2\pi$ (Hz)} & {$T_{1,q}$ (s)} & {$|\alpha|/2\pi$ (Hz)} & {$g/2\pi$ (Hz)} & {$\xi$} & {$C_{mq}$} \\ \hline
[0] & This work & 2026 &\multirow{3}{*}{LN-PNC}& \SI{3.60e9}{}& \SI{11e-3}{}& \SI{2.43e8}{}&  \SI{25}{}& 0.52-\SI{3.5e-3}{}& 1.3-\SI{8.6e5}{}& -- & -- & -- & -- & \SI{5.34e-4}{} & --\\ \relax
\cite{Wollack2022} & Wollack~et.~al. & 2022 &  & \SI{2.053e9}{} & \SI{1.23e-6}{}& \SI{1.59e+04}{}& -- & \SI{1.89e-1}{}& \SI{3.69e3}{}& \SI{2.443e9}{}& \SI{4.9e-6}{}& \SI{138e6}{}& \SI{9.5e6}{}& \SI{1.59e-2}{} & \SI{8.59e4}{}\\ \relax
\cite{Arrangoiz-Arriola2019} & Arrangoiz~et.~al. & 2019 & & \SI{2.405e9}{} & \SI{430e-9}{} & \SI{6.5e3}{} & -- & \SI{1.94e-1}{}& \SI{1.56e3}{}& \SI{2.417e9}{}& \SI{1.14e-6}{}& \SI{138e6}{}& \SI{15.7e6}{}& \SI{2.64e-2}{} & \SI{1.91e4}{}\\ \hline
\cite{Marti2024} & Marti~et.~al.  & 2024 & \multirow{3}{*}{AlN-HBAR} & \SI{5.023e9}{} & \SI{132e-6}{} & \SI{4.19e6}{}& --& \SI{1.55e-4}{}& \SI{6.5e2}{}& \SI{5.042e9}{}& \SI{16.9e-6}{}& \SI{185e6}{}& \SI{292e3}{}& \SI{2.97e-4}{} &\SI{3.02e4}{} \\ \relax
\cite{Franse2024} & Franse~et.~al.  & 2024 && \SI{6.064e9}{} &\SI{23.4e-6}{} &\SI{8.92e5}{} & --  &\SI{9.14e-5}{} &\SI{8.15e1}{} &\SI{6.067e9}{} &\SI{374e-9}{} &\SI{260e6}{} &\SI{197e3}{} &\SI{1.54e-4}{} &\SI{5.37e1}{} \\ \relax
\cite{Chu2017} & Chu~et.~al. & 2017 && \SI{6.678e9}{} & \SI{16.9e-6}{} & \SI{710e3}{} & -- & \SI{1.28e-4}{}& \SI{9.12e1}{}& \SI{6.684e9}{}& \SI{6e-6}{}& \SI{321e6}{}& \SI{260e3}{}& \SI{1.73e-4}{} &\SI{1.08e3}{} \\ \hline
\cite{Bienfait2019} & Bienfait~et.~al. & 2019 & \multirow{2}{*}{LN-SAW} & \SI{3.97e9}{} &\SI{ 1.2e-6}{} &\SI{2.99e4}{} & -- &\SI{1.59e-2}{} &\SI{4.83e2}{} &\SI{4.84e9}{} &\SI{7.6e-9}{} &\SI{179e6}{} &\SI{2.57e6}{} &\SI{2.7e-3}{}  &\SI{9.51}{} \\ \relax
\cite{Satzinger2018} & Satzinger~et.~al. & 2018 && \SI{3.985e9}{} & \SI{148e-9}{} & \SI{3.71e3}{}& -- & \SI{1.59e-2}{}& \SI{5.98e1}{}& \SI{4.725e9}{}& \SI{54e-9}{}& \SI{176e6}{}& \SI{7.3e6}{}& \SI{7.9e-3}{} &\SI{6.72e1}{} \\ \hline
\cite{OConnell2010} & O'Connell~et.~al.  & 2010 & AlN-FBAR & \SI{6.17e9}{} &  \SI{6.1e-9}{}&  \SI{2.36e2}{}&  -- &  \SI{1.2e-2}{}&  \SI{2.87}{}&  \SI{6.17e9}{}&  \SI{17e-9}{}&  \SI{19.7e6}{}&  \SI{67e6}{}&  \SI{1.77e-1}{}&  \SI{6.29e1}{} \\ \hline
\cite{Bozkurt2025} & Bozkurt~et.~al. & 2025 & \multirow{2}{*}{Si-ES} &\SI{4.918e9}{} &\SI{21e-3}{} &\SI{6.49e8}{} & -- &\SI{3.75e-6}{} &\SI{2.43e3}{} &\SI{5.107e9}{} &\SI{3e-6}{} &\SI{226e6}{} &\SI{200e3}{} &\SI{1.82e-4}{}  &\SI{9.95e4}{} \\ \relax
\cite{Bozkurt2023} & Bozkurt~et.~al. & 2023 && \SI{5.087e9}{} &\SI{265e-6}{} &\SI{8.47e6}{} &-- &\SI{5.92e-6}{} &\SI{5.01e1}{} &\SI{5.087e9}{} &\SI{306e-9}{}& -- &\SI{1.08e6}{}& -- &\SI{1.27e3}{}  \\ \hline
\cite{gerashchenko2025} & Gerashchenko~et.~al. & 2025 & SiN-ES & \SI{4.4e6}{} & \SI{5.9e-3}{} & \SI{1.63e5}{} & -- & \SI{1.6e-5}{} & \SI{2.65}{} & \SI{2.35e6}{} & \SI{22.4e-6}{} & \SI{418e6}{} & \SI{750}{} & \SI{1.79e-06}{} & \SI{1.17e1}{} \\ \hline
\cite{hu2025quartz} & Hu~et.~al. & 2025 & \multirow{2}{*}{Quartz-PNC} & \SI{97.2e6}{} & \SI{1.02e-3}{}& \SI{6.4e5}{} & -- & \SI{1.5e-4}{}& \SI{9.8e1}{}& -- & -- & -- & -- & \SI{4.2e-4}{} & -- \\
\cite{Emser2024} & Emser~et.~al. & 2024 &  & \SI{502.1e6}{} & \SI{51.0e-6}{}& \SI{1.61e5}{} & \SI{42.2}{} & \SI{1.6e-3}{}& \SI{2.55e2}{}& --& --& --& --& \SI{1.9e-3}{}& -- \\ \hline
\cite{Manenti2017} & Manenti~et.~al. & 2017 & \multirow{2}{*}{ Quartz-SAW} & \SI{523e6}{}&\SI{2.13e-6}{} &\SI{7e3}{} &\SI{402}{} &\SI{1.4e-3}{} &\SI{9.81}{} &\SI{2.6e9}{} &\SI{46e-9}{} &\SI{310e6}{} &\SI{5.7e6}{} &\SI{5.6e-3}{}  &\SI{5.03e2}{} \\ \relax
\cite{Manenti2016} & Manenti~et.~al. & 2016 && \SI{4.449e9}{}&\SI{1.23e-06}{} &\SI{34.5e3}{} &\SI{8.43e3}{} &\SI{1.4e-3}{} &\SI{4.83e1}{} &-- &-- &-- &-- &\SI{8.8e-3}{}  &-- \\ \hline
\cite{Andersson2022} & Andersson~et.~al. & 2022 & GaAs-SAW & \SI{3.87e9}{} &\SI{10.3e-6}{} &\SI{2.5e5}{} & \SI{38.7e3}{} &\SI{7e-4}{} &\SI{1.75e2}{} &-- &-- &-- &-- &\SI{2.03e-2}{} &-- \\ \hline
\end{tabular}}
\end{table*}

\section{Estimation of $k_{\text{eff}}^2$ for different mechanical resonators} \label{app:k2_estimation}
For some devices in the literature, $k_{\text{eff}}^2$ is explicitly quoted~\cite{Gustafsson2014,OConnell2010}, otherwise, we must estimate it from other provided parameters as detailed below.

The effective electromechanical coupling coefficient $k_{\text{eff}}^2$ is defined as the ratio between the amount of converted mechanical (electrical) energy $U_{\text{conv}}$, and the input electrical (mechanical) energy $U_{\text{input}}$, for a particular device.
Typically, $k_{\text{eff}}^2$ is defined within the context of piezoelectric resonators that have intrinsic piezoelectric coupling strengths quantified by $k^2=d^2/s^E\epsilon^\sigma$, where, for a particular mode, $d$ is the piezoelectric tensor component, $s$ is the compliance tensor component, $\epsilon$ is the dielectric constant, and is limited to $k_{\text{eff}}^2\leq k^2$. However, the definition of $k_{\text{eff}}^2$ can be extended to any device which converts energy between electrical and mechanical domains, and can be modeled by an equivalent circuit. 

The Butterworth-Van Dyke (BVD) equivalent circuit of a mechanical resonator is described by a static capacitance $C_0$ and motional capacitance (inductance) $C_m~(L_m)$ \cite{Wollack2021}. If we are far below the resonant frequency of the circuit and apply a voltage $V$, the motional inductance shorts and the energy of the circuit is partitioned between static $\frac{1}{2}C_0V^2$ and motional $\frac{1}{2}C_mV^2$ capacitances. From this, we can clearly express the ratio of converted mechanical energy to input energy in terms of the capacitances,
\begin{equation}\label{eq:keff2_from_BVD_capacitances}
    k_{\text{eff}}^2=\frac{U_{\text{conv}}}{U_{\text{input}}}=\frac{C_m}{C_m+C_0}.
\end{equation}
This equation can be used when equivalent circuit parameters are available~\cite{OConnell2010,Satzinger2018,Bienfait2019,Bozkurt2023,Bozkurt2025,hu2025quartz}. Otherwise Eq.~\ref{eq:keff2_from_BVD_capacitances} can be rewritten in terms of the characteristic resonance and anti-resonance frequencies,  $\omega_p^2=1/L_mC_m$ and $\omega_z^2=(C_m+C_0)/L_m C_m C_0$, that are obtained from the BVD circuit.
\begin{equation}\label{eq:keff2_from_Y_frequency_splitting}
    k_{\text{eff}}^2=\frac{\omega_z^2-\omega_p^2}{\omega_z^2}.
\end{equation}
For electromechanical devices which can be modeled and simulated with FEM, this equation can be used to obtain estimates of $k_{\text{eff}}^2$~\cite{Arrangoiz-Arriola2019,Wollack2022,Emser2024}. Additionally, Eq.~\ref{eq:keff2_from_Y_frequency_splitting} can be applied to electrostatically-biased mechanical membranes by recognizing that $\omega_z$ corresponds to the bare membrane frequency, while $\omega_p$ corresponds to the `softened' frequency under a bias~\cite{gerashchenko2025}.

For a high-overtone resonator, where the piezoelectric is a thin-film transducer of bulk excitations, the BVD circuit is modified to consist of many parallel motional capacitances $C_m=\sum_n C_{m,n}$, that when summed saturate the intrinsic coupling~\cite{Gokhale2026}.
\begin{equation}
    \sum_n C_{m,n} = C_0 \frac{k^2}{1-k^2}\approx C_0 k^2
\end{equation}
The number of modes over which the total motional capacitance is distributed is approximately $N\approx\omega_0/\omega_{\text{FSR}}$, therefore, each mode has a  diluted motional capacitance $C_{m,n}\approx\sum_n C_{m,n}/N$ and effective coupling:
\begin{equation}
    k_{\text{eff}}^2\approx\frac{C_{m,n}}{C_0} \approx k^2\frac{\omega_{\text{FSR}}}{\omega_0}
\end{equation}
This equation provides a generous upper bound on the effective electromechanical coupling coefficient for HBAR devices~\cite{Chu2017,Marti2024,Franse2024}.

In instances where there is insufficient information to estimate $k_\text{eff}^2$ from the reference, we use values from devices in the literature that are as comparable as possible. Specifically, \cite{Andersson2022} uses the value for a GaAs-SAW device from \cite{Gustafsson2014}, and \cite{Manenti2016,Manenti2017} uses a value for a Quartz-SAW device from \cite{Emser2022}.

\begin{figure}[h!]
\centering
\includegraphics[width=.4\columnwidth]{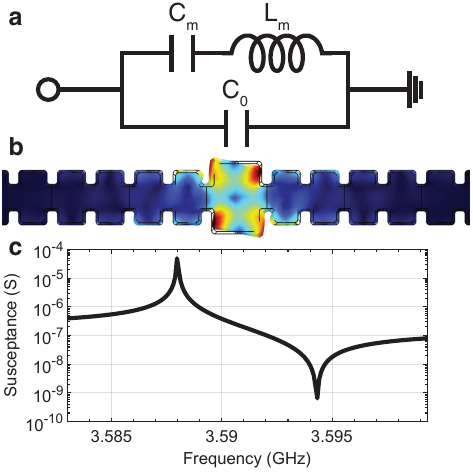}
\caption{\textbf{Equivalent circuit model and device simulation.} \textbf{a,} Butterworth-Van Dyke equivalent circuit model for a electromechanical resonator with motional capacitance (inductance) $C_m$ ($L_m$) and static capacitance $C_0$. \textbf{b,} Simulated displacement profile for the $\SI{3.6}{\GHz}$ device with $d=\SI{4760}{\nano\meter}$ presented in this work and represented in Tab.~\ref{tab:table_literature_parameter}. \textbf{c,} Simulated susceptance for the same device, exhibiting a pole and zero that corresponds to $k_{\text{eff}}^2=\SI{3.5e-3}{}$.}
\label{fig:SI_Fig_sim_Y_device}
\end{figure}

For the widest electrode gap spacing device ($d=\SI{4760}{\nano\meter}$) considered in this work, we use several methods to estimate $k_{\text{eff}}^2$.
For the first estimate, we use COMSOL Multiphysics to simulate the shear mode of interest, the displacement profile for this mode is shown in Fig.~\ref{fig:SI_Fig_sim_Y_device}b. From the admittance, we extract the pole and zero frequencies $\omega_p/2\pi=\SI{3.588}{\GHz}$ and $\omega_z/2\pi=\SI{3.5943}{\GHz}$ respectively, shown in Fig.~\ref{fig:SI_Fig_sim_Y_device}c, and use Eq.~\ref{eq:keff2_from_Y_frequency_splitting} to obtain an estimate of the effective coupling intrinsic to the device, $k_{\text{eff}}^2=\SI{3.5e-3}{}$.
We can also use the experimentally measured values of $\kappa_e/2\pi=\SI{25}{\Hz}$, $\omega_m/2\pi=\SI{3.6}{\GHz}$ and an estimate of the coupling capacitance $C_0=\SI{11.8}{\atto \farad}$ determined from a robust electrostatic capacitance simulation of the device electrode geometry. From these three parameters we calculate $k_{\text{eff}}^2= \kappa_e/(\omega_m^2C_0Z_0)=\SI{5.2e-4}{}$~\cite{Wollack2021}. This value represents a lower bound on $k_{\text{eff}}^2$ loaded by parasitics, such as additional stray capacitances stemming from the large device array, wirebonds, and packaging. These parasitics can be mitigated with improved device layout, packaging, and wirebonding.
To be conservative, we report both estimates in Fig.~\ref{fig:Fig_compare_performance},~\ref{fig:SI_M_T1_vs_year}, Tab.~\ref{tab:table_literature_parameter}.

\section{Shadow-Mask Fabrication and Metallization} \label{app:shadowmask_apparatus}

The nanofabrication described in the main text relies on custom tooling for shadow-mask fabrication, shadow-mask to substrate bonding and evaporation. This section describes our approach in detail. To improve accessibility and ease of replication, example wafer template CAD and 3D CAD models of all fixtures described here are provided in a \href{https://github.com/stanfordLINQS/nitride_shadowmask_metallization}{GitHub repository}.

We begin with a \qty{100}{\milli\meter} $\times$ \qty{525}{\micro\meter} FZ Si wafer with \qty{300}{\nano\meter} of LPCVD \ce{SiN$_{x}$} $(\le\qty{1000}{\mega\pascal}$) grown on both sides. We pattern the top side of the wafer with electron beam lithography (EBPG5200+, CSAR 6200.18). This design includes shadow-mask apertures (defining the nanodevice electrode patterns) and wafer-scale alignment marks. The pattern is transferred to the nitride with ICP RIE (10:1 \ce{CF4}:\ce{O2} plasma). The backside is then patterned with optical lithography (Heidelberg MLA 150, SPR3612) and an identical etch. In this step, backside alignment is used to align the release windows with the frontside aperture pattern.

Following the heated \ce{KOH} etch through the silicon (described in the main text), a \qty{700}{\nano\meter} \ce{Al} spacer layer is deposited around the shadow-mask apertures using electron beam evaporation (KJL LAB18). By depositing the spacer after the \ce{KOH} etch, we can use an arbitrary material for the spacer (Al will etch in KOH, for example). As it is not possible to spin resist on the released and patterned nitride membranes, this spacer deposition step is done with a coarse stencil mask. The mask is a standard laser-cut printed circuit board soldermask stencil of \qty{100}{\micro\meter} thickness. This stencil mask is aligned and secured $\approx\qty{100}{\micro\meter}$ from the shadow-mask wafer surface using a custom fixture, with which the Al layer is deposited. This spacer patterning approach natively avoids organics which may otherwise transfer to the LN during bonding.

The shadow-mask wafer contains 48 \qty[parse-numbers=false]{12\times7}{\milli\meter\squared} chips, each with a \qty[parse-numbers=false]{6.5\times4.5}{\milli\meter\squared} released nitride membrane. These are separated by \ce{KOH} etched trenches of width \qty{180}{\micro\meter} and depth $\approx\qty{100}{\micro\meter}$. Due to the fragility of the nitride membrane and the size of the wafer, hand cleaving of individual chips is challenging. For this reason a 3D printed jig with a living hinge cleaving arm was used to separate the chips, with a yield of $\approx\qty{90}{\percent}$. This concludes the shadow-mask fabrication process.

\begin{figure*}
	\centering
	\includegraphics[width=180mm]{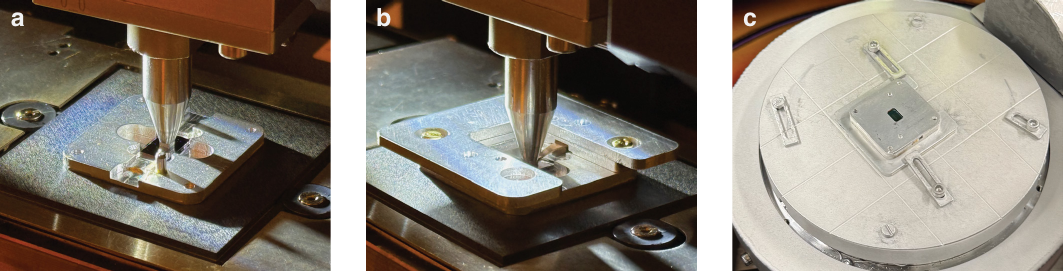}
	\caption{\textbf{Flip-chip bonding}. A shadow-mask chip and LN nanomechanics chip \textbf{a,} after aligned bonding, \textbf{b,} after assembly with clamping plates, within a Finetech Lambda 2 flip-chip bonder. Image \textbf{c,} illustrates the chip-fixture assembly ready for evaporation. Note that the fixture has been flipped to expose the shadow-mask between \textbf{b} and \textbf{c}.}
	\label{fig:S_flipchipbonding}
\end{figure*}

\begin{figure}[ht]
	\centering
	\includegraphics[width=0.5\columnwidth]{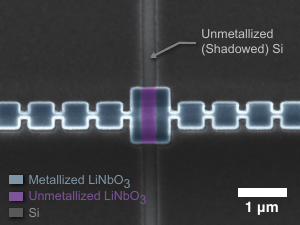}
	\caption{\textbf{Narrow shadowmasked electrode spacing.} Colorized SEM of on-defect contacts with \qty{360}{\nano\meter} separation metallized through a shadowmask. The grey background is the handle Si, which is metallized except where shadowed by the shadowmask bridge (as annotated) and phononic crystal.}
	\label{fig:S_smallspacingsem} 
\end{figure}

The flip-chip bonding fixture described in the main text is composed of three components, a baseplate that holds the shadow-mask chip and exposes the membrane for deposition, a midplate with clamping springs, and a top plate which mounts to the evaporator. The fixture supports $\qty{12}{\milli\meter}\times\qty{7}{\milli\meter}$ shadow-mask chips and $\qty{7.5}{\milli\meter}\times\qty{2}{\milli\meter}$ to $\qty{12}{\milli\meter}\times\qty{7}{\milli\meter}$ LN chips. A commercial flip-chip bonder (Finetech Lambda) is used to align the LN chip to the shadow-mask chip installed in the baseplate with a $\approx\qty{1}{\micro\meter}$ tolerance. The chips are compression bonded to \qty{10}{\newton}, and the midplate and top plates are installed to clamp the chips together (Fig. \ref{fig:S_flipchipbonding}). The midplate springs have sufficient clearance for a $\qty{2}{\milli\meter}\times\qty{2}{\milli\meter}$ die pick. At this point, the fixture and chip assembly is robust to vibration and shocks, and the spacing between the shadow-mask membrane and LN chip is roughly defined by the thickness of the spacer layer. We then deposit $\SI{50}{\nano\meter}$ of Al at normal incidence to form electrodes on the LN surface via electron beam evaporation (Plassys MEB550S, Fig. \ref{fig:S_smallspacingsem}). Following this, the top and midplate are removed, and the shadow-mask and LN chip are easily separated. This concludes the nanomechanical resonator fabrication.

\section{Shadow-Mask Distortion Is Negligible} \label{app:shadowmask_distortion}

\begin{figure*}
	\centering
	\includegraphics[width=180mm]{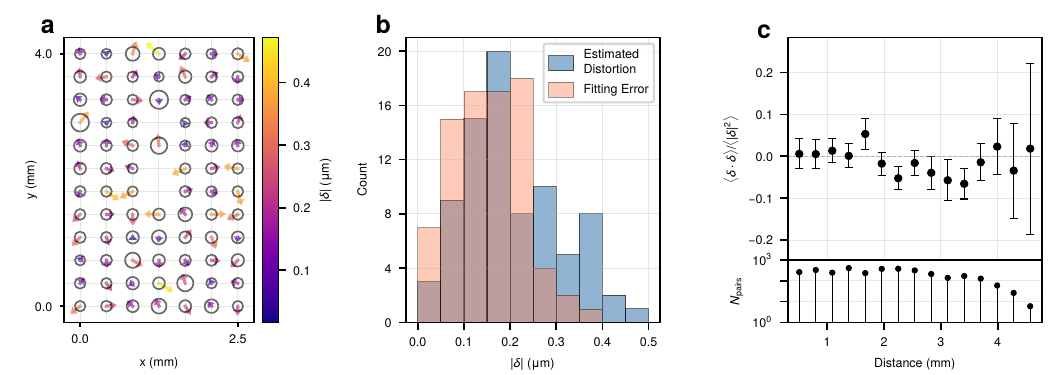}
	\caption{\textbf{Shadow-mask distortion characterization}. \textbf{a,} Distortion vectors measuring the displacement between each alignment marker and its corresponding shadow-mask deposited marker on the sample. Uncertainty is estimated through 30 independent fits on unique subsets of identified alignment marker boundary pixels, and represented by the circle radii. The $(x,y)$ position of each distortion arrow starting point and circle center shows the position of the marker pair on the chip. The empty regions are used for coarse alignment features. \textbf{b,} The distribution of measured distortion vector magnitudes is shown in blue, and is compared to the distribution expected from solely from fitting errors shown in red. \textbf{c,} The spatial autocorrelation of the magnitude of distortion. Each point shows the average autocorrelation across a \qty{300}{\micro\meter} wide bin of distances.}
	\label{fig:S_membranedistortion}
\end{figure*}

\newcommand*{\orh}[1]{\text{\overrightharp{\ensuremath{#1}}}}

There are several possible contributions to misalignment in shadow-masked patterning which are addressed in the above bonding scheme. The most direct contribution is the misalignment in the flip-chip bonding tool, which causes global translation and rotation of the observed pattern compared to the desired pattern. Here, we work within the $\approx\qty{1}{\micro\meter}$ translational tolerance of a research-grade flip-chip bonder. This is insufficient compared to the $\qty{100}{\nano\meter} - \qty{500}{\nano\meter}$ tolerance in alignment required for the PnC contact metallization. As a brute-force solution to this problem, we pattern a large array of $10\times6$ devices on each LN chip, offsetting the device and shadow-mask positions in a vernier capable of accommodating up to $\approx\qty{2}{\micro\meter}$ translational mask-pattern shift while still yielding aligned devices. The deliberate misalignment of the vernier limits the metallization yield, but this yield is now independent of the overall misalignment between the shadow-mask and LN chip. In practice, we see an expected and observed \qty{10}{\percent} to \qty{20}{\percent} metallization yield for devices with the narrowest electrode spacing, which increases to \qty{70}{\percent} to \qty{80}{\percent} for the devices with the largest electrode spacing. For these large electrode spacing devices, $\kappa_e$ is largely insensitive to electrode misalignment.

There are multiple routes towards improving the shadow-mask alignment. Production die bonders reach alignment tolerances of \SI{300}{\nm}, which would provide adequate alignment with no vernier for all but the smallest electrode spacing devices. Even for these devices, the vernier steps can be reduced, increasing the minimum yield. Beyond the bond tool tolerance, the shadow-mask technique admits self-alignment with $\le$ \SI{100}{\nm} tolerance \cite{tung_flip-chip_2014,bui_high-precision_2012}, using approaches such as etched divots in the shadowmask chip, compatible with our current fabrication, which mate with LN bumps. This may entirely decouple the ultimate shadowmask alignment from the bonder tolerance and provide sufficient alignment to remove the vernier entirely.

The second, more subtle, source of misalignment is due to deformation of the shadow-mask membrane relative to the chip. Asymmetric clamping forces can cause spacing variation over the membrane, which causes a position-dependent (nonlinear) shift of the deposited pattern \cite{tiggelaar_spreading_2007}. Furthermore, deposition of material onto a membrane modifies its strain, which can lead to a deflection of the membrane during deposition, `smearing' out a deposited pattern on the LN. This effect is exacerbated when the membrane is asymmetrically clamped at the beginning of the deposition. We use thick (\qty{525}{\micro\meter}) and therefore stiff silicon substrate and optimize the clamping uniformity to minimize this.

In general, the coordinates of a shadow-mask-deposited feature $\orh{x}_\text{actual}$ will vary from the desired position $\orh{x}_\text{target}$ by:

\begin{align}
    \orh{x}_\text{actual} = \mathbf{R} \orh{x}_\text{target} + \orh{\Delta x} + \orh{\delta}(\orh{x})
\end{align}

\noindent where $\mathbf{R}$ and $\orh{\Delta x}$ correspond to the global rotational and translational misalignment due to flip-chip bonding, and $\orh{\delta}(\orh{x})$ is an additional catch-all position-dependent misalignment that we name the `distortion'. There are many plausible contributions to $\orh{\delta}$. Here, our focus is on deformation of the shadow-mask membrane, which leads to a position-dependent displacement. Other contributions may include distortions in lithography, such as write-field stitching, or thickness variations in the spacer layer. While the deformation will fundamentally limit the alignment precision achievable with these shadowmasks, lithographic pattern or spacer thickness variations can be improved with better nanofabrication. It is difficult to identify the precise makeup of the net distortion, so measurements of $\orh{\delta}$ provide only an upper bound on the effect of shadow-mask deformation on the LN metallization.

We characterize $\orh{\delta}$ of our shadow-masking process by performing an aligned deposition of 81 markers over a \qty{2.5}{\milli\meter} $\times$ \qty{4}{\milli\meter} region of a test chip, forming a 7 x 12 index grid. We compare the position of the shadow-masked markers to their corresponding reference markers on the target chip, which are patterned through optical lithography, to measure $\orh{x}_\text{actual} - \orh{x}_\text{target}$ at each grid position. We least-squares fit the entire dataset to determine the global translation $\orh{\Delta x}$ and rotation $\mathbf{R}$ misalignments, where the residual of this fit for each grid position is $\orh{\delta}(\orh{x}_\text{target})$. Marker position and position uncertainty are estimated through computer vision on optical images of the chip following deposition, and uncertainties are propagated through Monte Carlo.

Fig. \ref{fig:S_membranedistortion} illustrates (a) the measured $\orh{\delta}$ at each grid position, and (b) the distribution of measured $|\orh{\delta}|$ compared to the distribution expected given measurement uncertainty and assuming $|\orh{\delta}|$ = 0. The experiment is generally limited by uncertainties but bounds distortion at $\le \qty{500}{\nano\meter}$ across the chip, well below the alignment error. We calculate the spatial autocorrelation

\begin{align}
    C(r) = \frac{1}{2n}\sum_{k}^n\frac{\left\langle \orh{\delta}(\orh{x_k}) \cdot \orh{\delta}(\orh{x'_k}) \right\rangle}{\left\langle |\orh{\delta} |^2\right\rangle}
\end{align}

\noindent where $k=1,2,3...n$ are the grid point pairs $(x_k,x'_k)$ that fulfill the condition $\left| |\orh{x_k} - \orh{x'_k}| - r \right| \le \Delta$. $2\Delta$ is our bin width that we set as \qty{300}{\micro\meter}. This autocorrelation does not reveal any length scale over which distortion is self-similar as would be expected for membrane deformation (Fig. \ref{fig:S_membranedistortion}c). The lack of strong evidence for distortion in this experiment is unsurprising, as deposition at normal incidence is to first order insensitive to height variation in the mask. More broadly, this suggests that shadow-mask membrane deformation is negligible for this fabrication process.

\section{Comparison of phonon lifetimes for several fabrication flows} \label{app:fab_variants}
\begin{figure*}[ht!]
	\centering
	\includegraphics{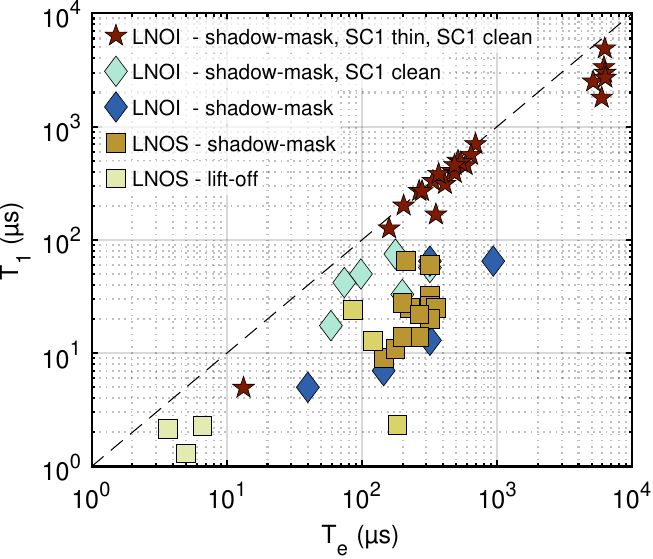}
	\caption{\textbf{Phonon lifetime versus external coupling lifetime for different fabrication flows}. Each point represents an individual resonator measured at large intracavity phonon number $\bar n\gg 1$. Measurements are grouped according to shared fabrication flows, with red, cyan, blue, orange and yellow data points corresponding to Tab.~\ref{tab:fab_variants} variants 1-3, 4, 5, 6, and 7-8 respectively. }
	\label{fig:fab_variants}
\end{figure*}

To support the reproducibility of the measured lifetimes and resolve the effect of the surface restoration techniques enabled by shadow-masking, we present an extended set of measurements that compare eight variants of the fabrication flow. 
Importantly, all devices share the same crystal orientation and shear mode defect resonance embedded within a phononic bandgap, while the precise geometries are perturbed slightly within a device array and between devices to allow for frequency multiplexing. As a result, the resonant frequency, $T_e$ and $T_1$ are naturally expected to vary between devices, therefore, we report $T_1/T_e$, a dimensionless coupling limit saturation ratio, as the informative statistic to compare between designs. 
The main fabrication steps as well as the mean, standard deviations, and sample sizes of the $T_1/T_e$ ratios are reported in Table~\ref{tab:fab_variants}. Coming from the same lithium niobate on insulator (LNOI) wafer, variants 1-3 correspond to the devices discussed in the main text and variants 4-6 represent the switch to shadow-masking and experimentation with different surface restoration techniques. Meanwhile, variants 7-8 represent the legacy fabrication process with lift-off metalization and a lithium niobate on silicon (LNOS) stack. 
We find that in the absence of metal losses, the implementation of shadow-mask metalization and the surface restoration techniques it enables (variants 1 and 2) results in higher $T_1/T_e$, indicating reduced intrinsic phonon losses compared to the other fabrication variants.
Fig.~\ref{fig:fab_variants} shows $T_1$ and $T_e$ for all resonators that enter this comparison. 

\begin{table}[ht!]
\centering
\caption{Process variants and statistics. Each variant lists the film
stack, electrode gap, metalization method, whether SC1 thinning or final SC1 cleaning were performed, the mean and standard deviation of the
external-coupling saturation $T_1/T_e$, and the number
of measured resonators $n$ within the variant. The devices presented in the main text belong to variants 1, 2 and 3. $^\dagger$Variant 3 has an insufficient sample size to report a standard deviation.}
\label{tab:fab_variants}
\begin{ruledtabular}
\begin{tabular}{l l c c c c l c}
Variant & Stack &Electrode gap& Metalization & SC1 thin & SC1 clean & $T_1/T_e$ & $n$  \\
\hline
1& LNOI  & \SI{4760}{\nm} & shadow-mask & Y & Y &  \SI{0.50\pm.16}{} &  6  \\
2& LNOI  & \SI{2800}{\nm} & shadow-mask & Y & Y &  \SI{0.86\pm.15}{} &  16  \\
3& LNOI  & \SI{360}{\nm} & shadow-mask  & Y & Y &  \SI{0.37}{}$^\dagger$ &  1  \\
4& LNOI   & \SI{1700}{\nm} & shadow-mask & N & Y & \SI{0.36\pm.17}{} & 6   \\
5& LNOI   & \SI{1700}{\nm} & shadow-mask & N & N & \SI{0.09\pm0.06}{} & 6   \\
6& LNOS   & \SI{1850}{\nm} & shadow-mask & N & Y & \SI{0.11\pm.07}{} & 13   \\
7& LNOS   &  \SI{1700}{\nm} & liftoff-resist & N & N & \SI{0.13\pm.14}{} & 3   \\
8& LNOS   &  \SI{400}{\nm} & liftoff-resist & N & N & \SI{0.42\pm.16}{} & 3   \\
\end{tabular}
\end{ruledtabular}
\end{table}

\end{document}